\documentclass[twocolumn,reprint,amsmath,amssymb,aps,prl,floatfix,longbibliography,]{revtex4-2}

\usepackage{siunitx}
\usepackage{multirow}
\usepackage[pagewise]{lineno}
\usepackage{graphicx}
\usepackage{dcolumn}
\usepackage{bm}
\usepackage{titlesec}
\usepackage{capt-of} 
\usepackage{upgreek}
\usepackage{xcolor}
\usepackage{physics}
\usepackage{natbib}
\usepackage{indentfirst}
\titlespacing*{\section}{0pt}{1.2\baselineskip}{1\baselineskip}
\begin{document}


\title{High-Efficiency Plasma-Based Compressor for Ultrafast Soft X-ray Free-Electron Lasers}

 \author{Mingchang Wang \textsuperscript{1,2} }\thanks{These authors contributed equally to this work.}
 \author{Li Zeng \textsuperscript{3} }\thanks{These authors contributed equally to this work.}
 \author{Bingbing Zhang \textsuperscript{3}}
  \author{Qinghao Zhu \textsuperscript{3}}
 \author{Xiaozhe Shen \textsuperscript{3}}
 \author{Xiaofan Wang \textsuperscript{3}}\email{wangxf@mail.iasf.ac.cn}
 \author{Qinming Li \textsuperscript{3}}\email{liqinming@mail.iasf.ac.cn}
 \author{Weiqing Zhang\textsuperscript{1}}\email{weiqingzhang@dicp.ac.cn}
 \affiliation{
        \textsuperscript{1}Dalian Coherent Light Source and State Key Laboratory of Chemical Reaction Dynamics, Dalian Institute of Chemical Physics, Chinese Academy of Sciences, Dalian, 116023, China\\
         \textsuperscript{2}University of Chinese Academy of Sciences, Beijing, 100049, China\\
 		\textsuperscript{3}Institute of Advanced Light Source Facilities, Shenzhen 518107, China}


\begin{abstract}
The generation of intense, femtosecond-scale X-ray pulses is crucial for probing matter under extreme temporal and field conditions. Current chirped-pulse amplification (CPA) techniques in free-electron lasers (FELs), however, face efficiency limitations in the soft X-ray regime due to the inherent constraints of conventional optical compressors. To address this challenge, we propose a high-efficiency plasma-based compressor utilizing highly ionized noble gas plasma. Exploiting strong refractive index dispersion near ionic resonances, this scheme achieves over 70\% transmission efficiency around \SI{5.2}{nm}, and is extendable to other highly charged ions for operation across the soft X-ray to vacuum ultraviolet range. Simulations demonstrate that a \SI{25}{fs} FEL pulse can be compressed to \SI{1.4}{fs} with peak power boosted to over \SI{100}{GW}, while maintaining high energy throughput. This approach overcomes the long-standing efficiency bottleneck of soft X-ray CPA and opens a scalable path toward compact, high-brightness attosecond FEL sources.
\end{abstract}

\maketitle

Free-electron lasers (FELs) operating in the attosecond-to-femtosecond regime, with their unprecedented temporal resolution and ultrahigh peak power, have revolutionized the study of ultrafast dynamics in matter~\cite{emma2010first,duris2020tunable}. Soft X-ray ultrashort pulses, in particular, enable direct real-time observation of electron transitions~\cite{driver2024attosecond,li2022attosecond} and transient quantum phenomena~\cite{barillot2021correlation}, overcoming the temporal limitations of conventional spectroscopy. Furthermore, such high-power outputs are indispensable for applications requiring strong-field excitation~\cite{boll2022x} or quantum control~\cite{richter2024strong}, underscoring the critical role of intense unltrafast FELs in advancing attosecond science.

To reach the femtosecond-to-attosecond regime, various approaches have been proposed, such as emittance-spoiling~\cite{Ding2015Generating}, low-charge operation~\cite{Huang2017Generating}, and so on~\cite{Zholents2005Method,  Prat2015Simple, Guetg2018Generation, tanaka2015proposal, maroju2020attosecond, Mirian2021Generation}. However, these methods often reduce the effective charge participating in the amplification process, thereby limiting the attainable pulse energy and peak power. In contrast, the application of chirped pulse amplification (CPA) to FELs offers a more efficient route for generating intense femtosecond FEL pulses~\cite{gauthier2016chirped, Li2022Femtosecond}, enabling efficient energy extraction from the whole electron bunch.

A fundamental limitation in CPA-based FELs arises during the recompression stage. Conventional grating-based compressors exhibit low efficiency at X-ray wavelengths, with typical transmission efficiencies below 5\%~\cite{gauthier2016chirped, duris2020tunable}. While multilayer mirrors and Bragg crystals provide partial mitigation within specific spectral bands~\cite{shvyd2006x, bourassin2015multilayer, Li2022Femtosecond}, a universal high-efficiency compression method remains elusive across the broad 1--10 nm regime. This bottleneck severely restricts the performance and scalability of soft X-ray CPA schemes.

To address this challenge, we propose a fundamentally different approach: a plasma-based compressor utilizing noble gas ions. By exploiting strong dispersion near resonant transitions in highly ionized atoms, this method offers broadband group velocity control and high transmission in the soft X-ray regime. Specifically, we demonstrate that a plasma column can compress a \SI{25}{fs} FEL pulse to \SI{1.4}{fs} with peak power exceeding \SI{100}{GW}, while maintaining a transmission efficiency of over 70\%, a dramatic improvement over existing optical methods. Unlike conventional methods, this plasma compressor is scalable, tunable, and compatible with existing FEL infrastructures, offering a practical and transformative route to bridge the soft X-ray CPA gap and enabling next-generation high-brightness attosecond sources.

\begin{figure*}[hbt!]
    \centering
    \includegraphics[width=0.95\textwidth]{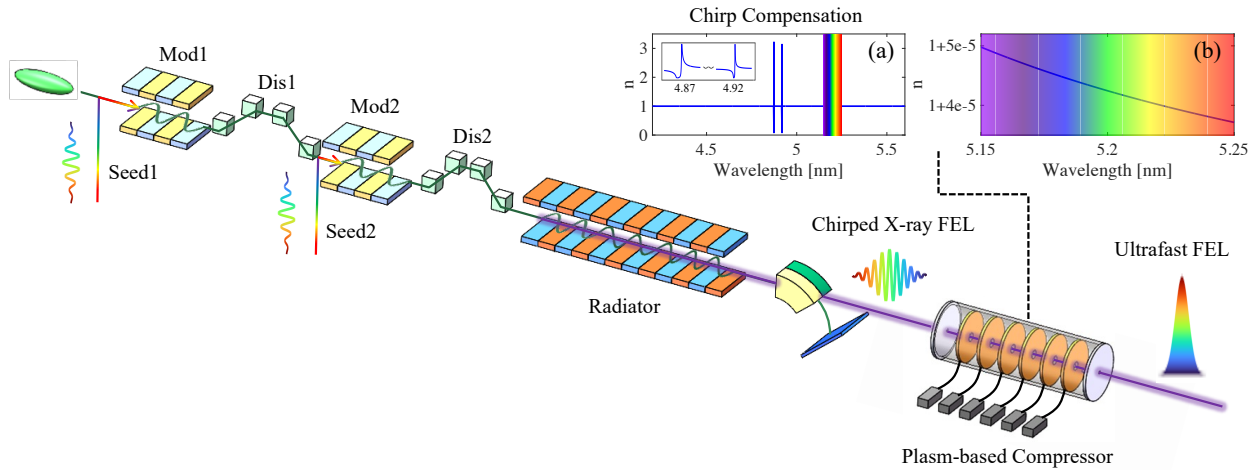}  
    \caption{Schematic diagram of chirped pulse amplification based on a noble gas plasma compression device. (a) Calculated real part $n$ of the refractive index for Ar$^{8+}$ in the 4.2–5.6~nm range, showing strong resonances near 4.87 and 4.92~nm. (b) Fine-scale variation of $n$ within the 5.15–5.25~nm compression window, illustrating a nearly linear slope that closely matches the chirp profile of the EEHG-generated FEL pulse.}
    \label{fig1}
\end{figure*}

Figure~\ref{fig1} illustrates the schematic layout of the proposed scheme, which integrates a noble-gas plasma compressor with an echo-enabled harmonic generation (EEHG) FEL~\cite{PhysRevLett.102.074801, PhysRevSTAB.12.030702}. EEHG is a well-established external seeding technique that enables high-harmonic up-conversion of long-wavelength seed lasers into fully coherent soft X-ray radiation~\cite{rebernik2019coherent}. Its ability to imprint and preserve frequency chirp from broadband seed lasers makes it highly compatible with CPA strategies in soft X-ray FELs. In our configuration, two frequency-chirped seed lasers initiate the EEHG process, imprinting chirp characteristics onto the high-harmonic microbunching and subsequently onto the FEL radiation. Specifically, we operate at the 53rd harmonic of the seed laser, generating soft X-ray pulses centered near 5.2~nm with a positive chirp inherited from the seed. These chirped pulses are then directed into a plasma cavity, where strong dispersion near atomic resonances enables efficient group-delay compensation. This two-stage approach, where EEHG enable coherent frequency multiplication and chirp preservation, and the plasma compressor provides efficient pulse shortening, yields ultrafast high-power FEL output.

To ensure sufficient dispersion in the soft X-ray regime, we employe a 1-meter-long plasma cavity filled with highly charged argon ions Ar$^{8+}$ at a density of $2 \times 10^{21}$cm$^{-3}$. This plasma is generated via a gas-puff Z-pinch discharge, which offers a compact and reproducible source of generating high-density noble gas plasma~\cite{bleach1983extreme, shumlak2001evidence, pikuz2002high}. Compared to other plasma generation methods, such as electron cyclotron resonance~\cite{chung2002electron}, inertial confinement~\cite{glenzer2010symmetric,betti2016inertial}, laser ablation~\cite{irimiciuc2020target}, Z-pinch systems are well suited for generating extended volumes of highly charged ions. These systems have demonstrated total electron densities exceeding $5 \times 10^{21}$cm$^{-3}$~\cite{bleach1983extreme}, making them a practical choice for our application. In our implementation, segmented electrodes and modular high-voltage control are assumed to ensure discharge uniformity and stability.

The key to our compression strategy lies in the strong dispersion near atomic resonances of Ar$^{8+}$ ions. Near resonance, the complex refractive index $\tilde{n}(\omega) = n(\omega) + i\beta(\omega)$ follows the Lorentz–Lorenz model~\cite{drescher2018extreme}:
\begin{equation}
\frac{\tilde{n}^2 - 1}{\tilde{n}^2 + 2} = \frac{N e^2}{3 m_e \epsilon_0} \sum_j \frac{f_j}{\omega_{0j}^2 - \omega^2 - i\Gamma_j \omega},
\label{eq:lorentz}
\end{equation}
where $N$ is the ion density, $f_j$ the oscillator strength, and $\omega_{0j}, \Gamma_j$ the resonance frequency and width, respectively. Figure~\ref{fig1}(a) shows the calculated real part $n(\omega)$ of the refractive index for Ar$^{8+}$ in the 4.2–5.6 nm range, based on this model and resonance data from the NIST database~\cite{Nistdatabase}. Two prominent peaks near 4.87 and 4.92~nm correspond to different fine-structure components of the 2s$^2$2p$^6$ $\rightarrow$ 2s$^2$2p$^5$3s transitions, and they induce strong anomalous dispersion and absorption.

Zooming in on the working window, Fig.~\ref{fig1}(b) highlights the behavior of $n$ within 5.15–5.25~nm (shaded in rainbow). In this region, the refractive index satisfies $n > 1$ with a negative slope—i.e., normal dispersion, where longer wavelengths travel faster than shorter ones. This trend is essential for compensating the positive chirp of the FEL pulse. In contrast, the $n < 1$ region below resonance provides flat group velocity and fails to compress pulses effectively. Thus, selecting a bandwidth slightly above the resonance ensures both effective dispersion and manageable absorption.

To quantify the compressibility, the wavevector $k(\omega) = n(\omega)\omega / c$ is expanded as a Taylor series. The second-order coefficient $k_2$ captures the group delay dispersion, which governs the main compression effect.
For a linearly chirped input pulse described by $\omega(t) = \omega_0 + \frac{a}{2\tau_0^2}t$, where $a$ is a dimensionless coefficient indicating the strength of the longitudinal chirp, optimal compression occurs when the chirp-induced quadratic phase is precisely canceled by the plasma's dispersion. This happens under the condition
\begin{equation}
\frac{a}{2} = \frac{1}{2} k_2 \frac{a^2}{4\tau_0^2} Z,
\end{equation}
with $Z = 1$~m being the interaction length, yielding a transform-limited pulse duration $\tau_0$. 
The spectral slope of $n(\omega)$ in the selected window is closely matched to the frequency-time slope of the positively chirped FEL pulse generated via EEHG, enabling efficient plasma-based chirp compensation and the production of high-power, ultrashort FEL pulses.

\begin{figure}[htb!]
    \centering
    \includegraphics[width=\columnwidth]{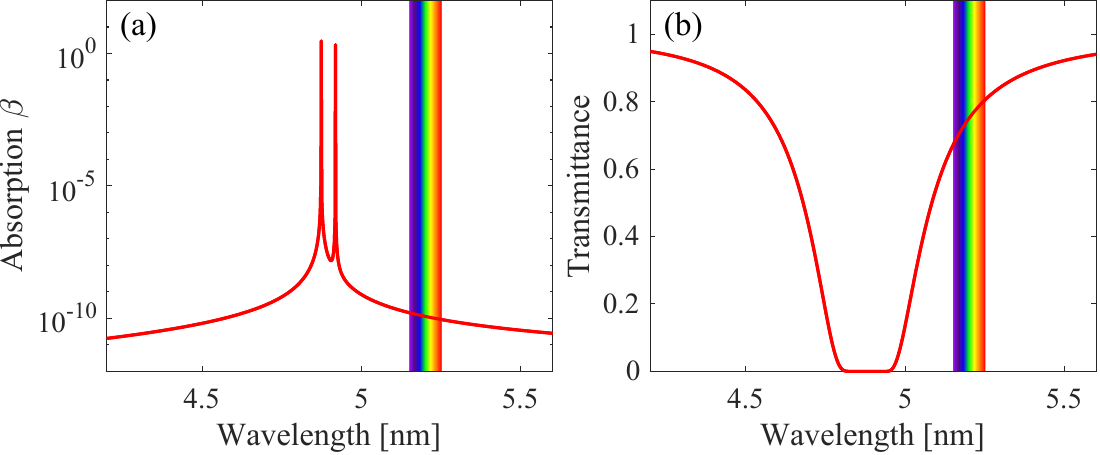}
    \caption{Absorption coefficient $\beta$ (a) and transmission efficiency (b) near the Ar$^{8+}$ resonance, both plotted over the 4.2–5.6 nm wavelength range.}
    \label{fig2}
\end{figure}

Figure~\ref{fig2}(a) presents the imaginary part $\beta$ of the refractive index, which governs absorption. Strong absorption occurs near the resonances at 4.87 and 4.92 nm, where $\beta$ peaks and transmission approaches zero. In contrast, in the selected 5.15–5.25 nm region, $\beta$ is significantly reduced. Figure~\ref{fig2}(b) shows the corresponding transmission efficiency, calculated via the Beer–Lambert law:
\begin{equation}
T = \exp\left(-\frac{4\pi \beta Z}{\lambda}\right),
\label{eq:transmission}
\end{equation}
The transmission remains in the range of 60–80\%, making the plasma compressor both effective and practical for high-flux applications.

Other plasma components, such as free electrons and low-charge-state ions, are negligible in this context. The plasma frequency of free electrons lies in the UV range ($\sim$265~nm) and has no appreciable effect in the soft X-ray domain. Meanwhile, the proportion of other ion valence states such as Ar$^{7+}$ is suppressed at suitable electron temperature, and their resonances lie far outside the compression band (see Supplementary Material). Ar$^{8+}$ thus dominates both dispersion and transmission in the target window, enabling efficient chirp compensation while preserving pulse energy.

To evaluate the performance of the proposed plasma-based compression scheme, we conducted comprehensive numerical simulations using realistic parameters for a high-gain seeded FEL. The simulations were performed with both \textsc{GENESIS} 1.3 \cite{Reiche1999GENESIS} and \textsc{OCELOT}. The electron beam was configured with an energy of 2.5 GeV, a relative energy spread of $1 \times 10^{-4}$, a normalized emittance of 0.5 mm·mrad, and a peak current of 1.5 kA. The root-mean-square (RMS) transverse beam size was approximately 35 $\upmu$m.

The seed laser consisted of two pulses at 275.6 nm, each with an intrinsic spectral bandwidth of approximately 6.5\%, enabling  linear temporal stretching from a transform-limited duration of $\sim$6.5 fs to 125 fs (FWHM). With a waist size of approximately 0.5 mm and a peak power of about 130 MW, these chirped pulses interacted with the electron beam in two modulators, each 2 m long with a 9 cm period length, imprinting energy modulation that faithfully inherited the seed chirp. The modulations were converted into high-harmonic density structure via two dispersion sections with momentum compaction factors of 2.60 mm and 0.10 mm, respectively. The bunched beam subsequently entered a radiator composed of four 4-meter-long undulators with 43 mm period length, tuned to 5.2 nm to generate and amplify chirped FEL pulses that matched the Ar$^{8+}$ plasma resonance window.

\begin{figure}[htb!]
\centering
\includegraphics[width=\columnwidth]{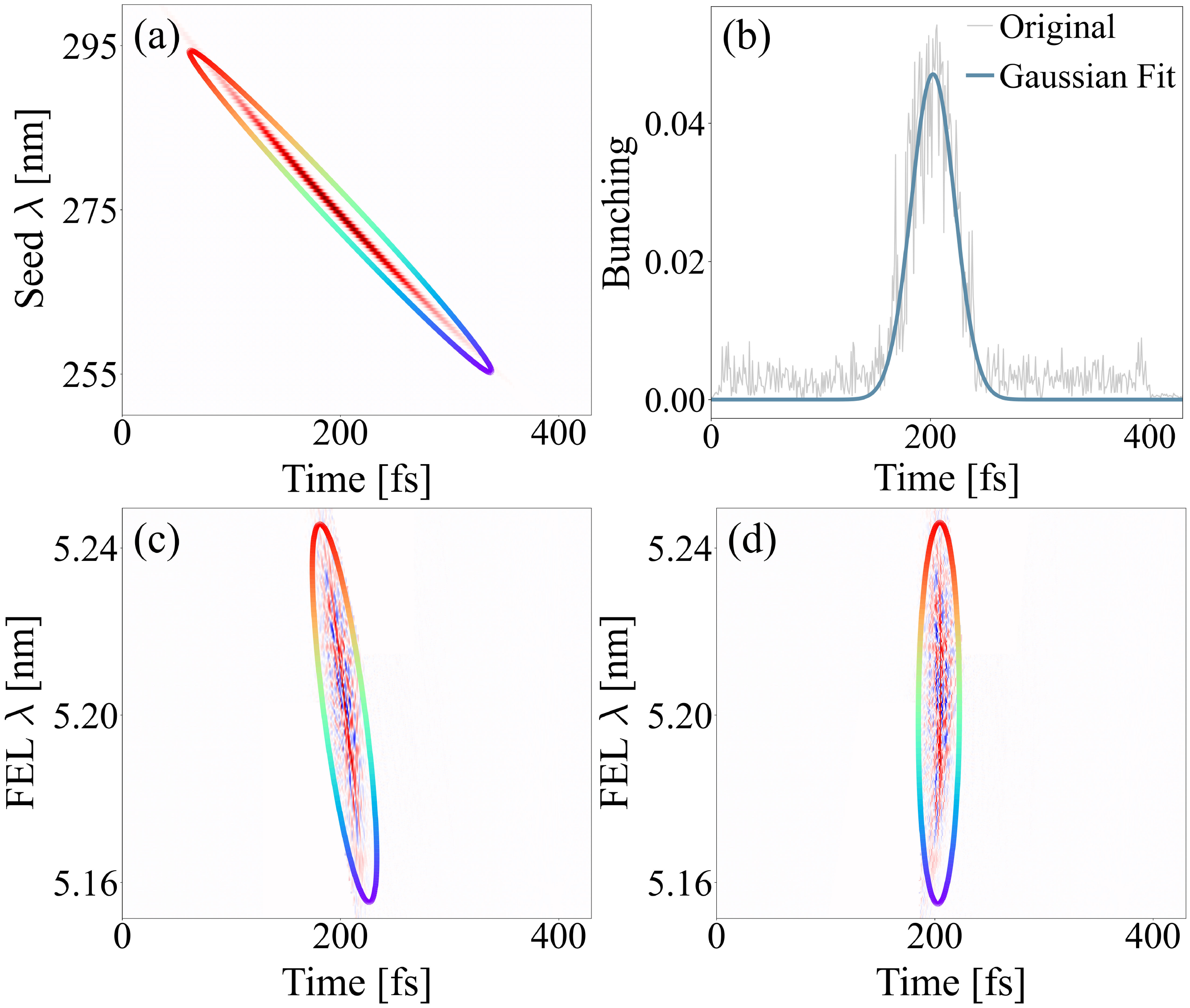}
\caption{Wigner distributions of the seed laser pulse (a), and FEL pulses before (c) and after (d) plasma-based compression. Panel (b) shows the bunching factor of the electron beam at entrance of the radiator.}
\label{fig3}
\end{figure}

To characterize the temporal and spectral properties of the FEL pulses, we employed the Wigner distribution, which privides a joint time–frequency representation that simultaneously reveals the evolution of chirp and spectral structure. Figure \ref{fig3}(a) shows the Wigner distribution of the seed pulse, clearly revealing a linear chirp—high-frequency components lead, while low-frequency components lag—across a duration of 125 fs.
At the radiator entrance, the 53rd-harmonic bunching factor [Fig.~\ref{fig3}(b)] reaches a peak of $\sim$5\% with a temporal width of 46.7 fs, notably shorter than the seed pulse.
This reflects an intrinsic feature of EEHG, wherein the microbunching length is inherently reduced due to the localized phase space modulation, with higher harmonic orders yielding progressively narrower temporal structures. The Wigner distribution of the FEL pulse prior to compression is shown in Fig.~\ref{fig3}(c), demonstrating that the linear chirp is well preserved through the FEL amplification process. 

After propagation through the Ar$^{8+}$ plasma, the pulse undergoes substantial temporal compression [Fig.~\ref{fig3}(d)]. This behavior arises from the anomalous group velocity dispersion near the plasma resonance: long-wavelength components at the pulse rear propagate faster than short-wavelength components at the front, thereby enabling temporal overlap and compression. While the dispersion profile exhibits smooth curvature rather than strict linearity, its local slope is broadly compatible with the linear chirp of the seed laser, which was deliberately chosen to represent a general, non-tailored case. This partial mismatch leads to residual phase distortions that hinder full energy convergence at the main peak, resulting in side lobes and limiting the effective peak power enhancement.

\begin{figure}[htb!]
\centering
\includegraphics[width=0.85\columnwidth]{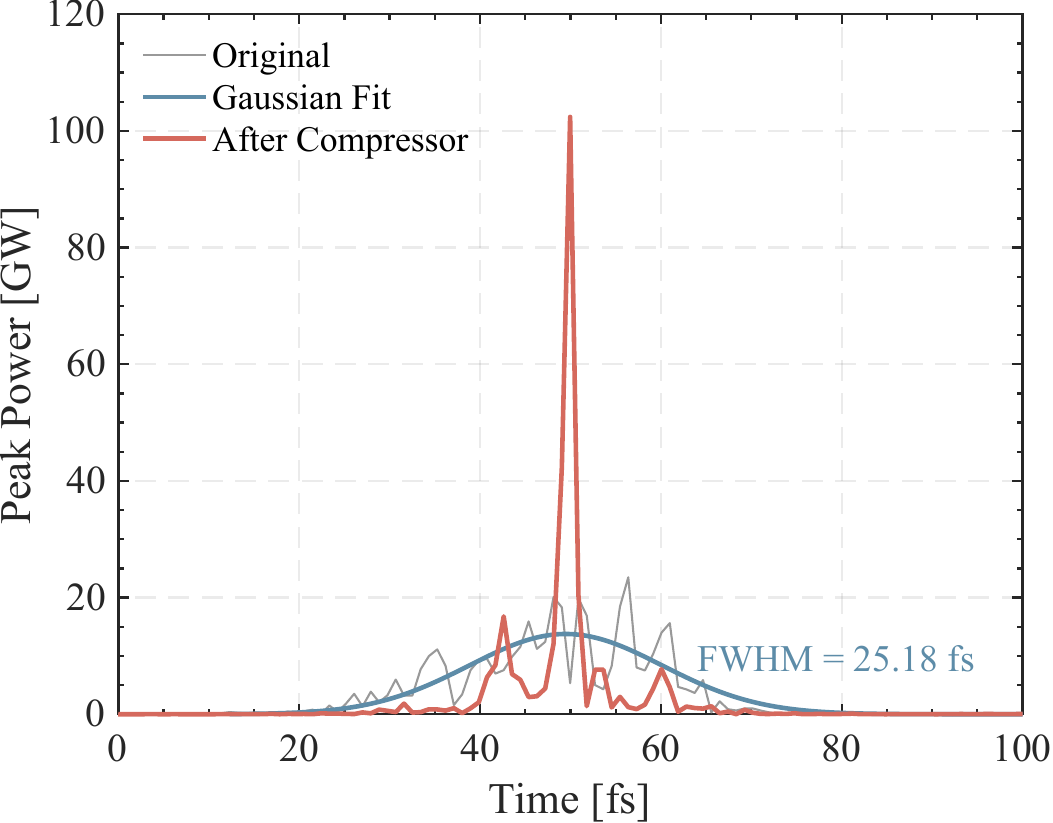}
\caption{Temporal power profiles of FEL pulses before (blue) and after (red) plasma-based compression.}
\label{fig4}
\end{figure}

The temporal power profiles of the FEL pulse before and after compression are shown in Fig.~\ref{fig4}. According to standard CPA-free FEL pulse scaling, the expected pulse duration can be estimated as $\sigma_t^{\text{FEL}} \approx h^{-\frac{1}{3}} \sigma_t^{\text{seed}} \approx 33.27$\,fs~\cite{gauthier2016chirped, Finetti2017Pulse}, where $h$ is the harmonic number and $\sigma_t^{\text{seed}}$ is the seed pulse duration. In our simulation, however, the uncompressed FEL pulse exhibits a slightly shorter duration of 25.18\,fs (FWHM). This discrepancy arises because the seed laser’s spectral bandwidth exceeds the gain bandwidth of the modulator, leading to selective amplification of central spectral components. This spectral filtering effect narrows the temporal region of effective energy modulation, thereby reducing the resultant FEL pulse duration.

After passing through the plasma compressor, the FEL pulse is significantly shortened to 1.36 fs (FWHM), accompanied by an increase in peak power from 23.5 GW to 102.4 GW. The pulse energy decreases from 368 $\upmu$J to 271 $\upmu$J, corresponding to a transmission efficiency of 73.6\%. This value aligns well with the spectral transmission shown in Fig.~2(b), where the selected working window (5.15–5.25 nm) lies moderately above the Ar$^{8+}$ resonance (4.92 nm), avoiding strong absorption while still providing sufficient dispersion. Notably, the reported pulse duration refers only to the main peak, excluding residual side lobes that originate from higher-order dispersion and carry a non-negligible fraction of the energy. These features limit the peak power enhancement achievable under linear-chirp seeding.

Nevertheless, the compression performance achieved with a non-tailored seed underscores the robustness and generality of the proposed scheme. Further performance gains are anticipated through the introduction of a pre-compensating nonlinear chirp in the seed laser, made feasible by the chirp-preserving nature of seeded FELs. Such refinement would allow more precise phase matching with the plasma dispersion and support near-transform-limited pulse formation.

\begin{figure}[htb!]
\centering
\includegraphics[width=\columnwidth]{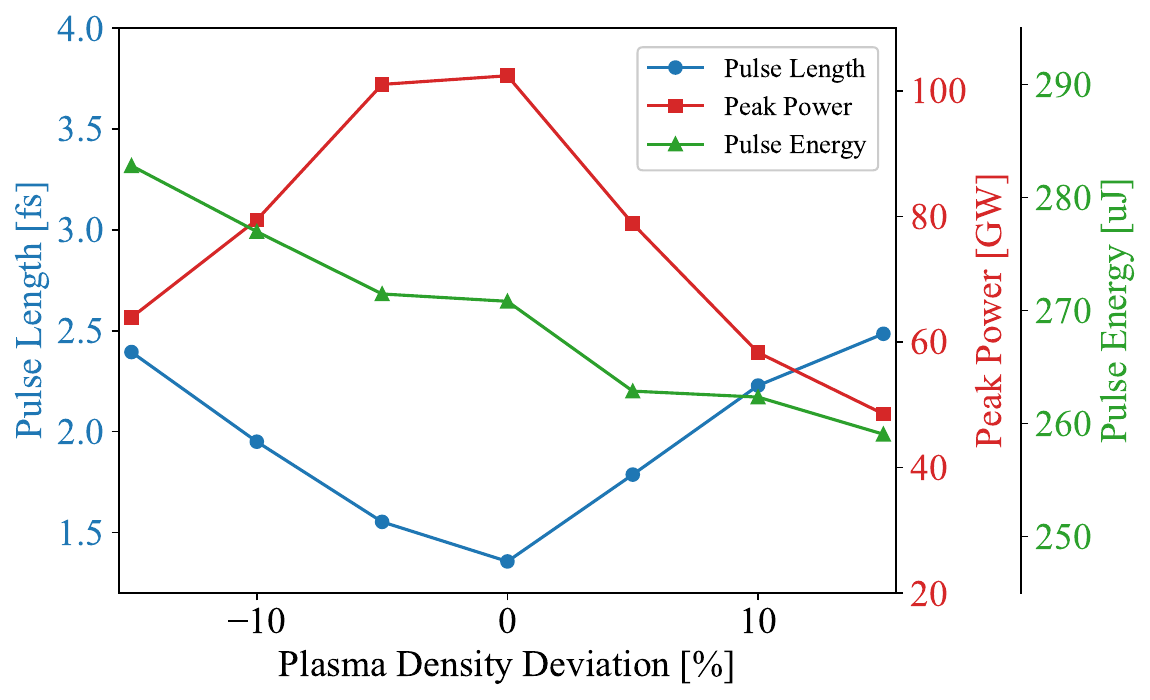}
\caption{Impact of plasma density on pulse duration (blue), peak power (red) and pulse energy (yellow).}
\label{fig5}
\end{figure}

The transverse properties of the FEL pulse at saturation also satisfy practical requirements: the RMS beam size is 101.9 $\upmu$m and the divergence is 16.0 $\upmu$rad. These values fall well within the tolerances imposed by plasma uniformity and beamline apertures, ensuring that the transverse phase distortion remains minimal during propagation. Maintaining a compact transverse dimension is essential—not only to preserve compression fidelity by avoiding spatial inhomogeneity in the plasma refractive index, but also to enable tighter plasma chamber apertures, which facilitate vacuum isolation and beamline integration. The simulated beam parameters validate the practical feasibility of implementing this scheme in realistic experimental settings.

We further investigated the robustness of the scheme with respect to plasma density variations around the optimal value of $2 \times 10^{21}~\text{cm}^{-3}$. As shown in Fig.~\ref{fig5}, deviations from this optimal density—whether positive or negative—inevitably shift the system away from the ideal compression condition, resulting in longer pulse durations and lower peak powers. These two parameters exhibit a symmetric and monotonic trend: as the density increases or decreases, the pulse stretches and the peak power diminishes. In contrast, the pulse energy exhibits a distinctly different, asymmetric behavior. It peaks at the lowest plasma density, where absorption is minimal, and gradually decreases with increasing density due to stronger plasma absorption. Despite this, even at a $+15\%$ density deviation—where absorption is strongest—the output pulse energy remains as high as 259~$\upmu$J, corresponding to a transmission efficiency of 70.4\%. Importantly, despite these variations, the pulse duration remains below 3 fs across the entire $\pm 15\%$ density range, and the peak power stays above 48 GW. This divergence in trends highlights the distinct physical mechanisms underlying temporal broadening and absorption, and demonstrates the inherent resilience of the scheme under experimentally relevant plasma conditions.

To evaluate the scheme's scalability, we extended the analysis to a more significant, tenfold reduction in plasma density, down to $2 \times 10^{20}~\text{cm}^{-3}$. The dispersion curve retains its overall shape, with both slope and curvature reduced proportionally. This indicates that the dispersion characteristics are primarily governed by spectral proximity to resonance, rather than absolute density. Accordingly, to achieve a comparable dispersion slope at lower density, the operating band must be shifted closer to the resonance (e.g., from 5.15–5.25 nm to 5.0–5.1 nm). Notably, transmission remains high within this shifted spectral window at reduced density, alleviating absorption concerns. Given that the seed laser chirp can be flexibly pre-adjusted in experiments, strong compression performance is thus attainable over a wide density range, confirming the scheme's feasibility under less stringent plasma conditions.

This work proposed and validated a high-efficiency chirped pulse compression scheme for soft X-ray free-electron lasers, based on the resonance-enhanced dispersion of noble gas plasmas. By exploiting the strong frequency dependence of the real part of the refractive index near ionic resonances, this method enables linear chirp compensation with transmission efficiencies exceeding 70\%. Numerical simulations demonstrate that a 25 fs FEL pulse can be compressed to approximately 1.4 fs, with the peak power enhanced beyond 100 GW, while preserving more than 70\% of the pulse energy.

This approach represents a transformative advancement in chirped-pulse amplification for FELs, overcoming the intrinsic efficiency limitations of grating-based compressors in the soft X-ray regime. Moreover, by tailoring the plasma species and charge states, the scheme can be extended across the soft X-ray to VUV spectral range (see Supplementary Material). The proposed concept opens a scalable pathway to sub-femtosecond, high-peak-power soft X-ray pulses, offering new opportunities for attosecond science, nonlinear X-ray optics, and photon–matter interaction studies in previously inaccessible regimes.

\begin{acknowledgments}
This work is supported by the Scientific Instrument Developing Project of Chinese Academy of Sciences (Grant No. GJJSTD20220001), the National Natural Science Foundation of China (Grant Nos. 12305359, 22288201, 22303055), the LiaoNing Revitalization Talents Program (Grant No. XLYC2202030), the Strategic Priority Research Program of the Chinese Academy of Sciences (Grant No. XDB0970000, subject No. XDB0970100) and the Shenzhen Science and Technology Program (Grant No. RCBS20221008093327057) . We acknowledge the support of the staff members of Dalian Coherent Light Source (https://cstr.cn/31127.02.DCLS). We also sincerely thank Dr. Carlo Callegari and Dr. Kevin C. Prince of Elettra-Sincrotrone Trieste for their insightful discussions and valuable contributions to this work.
\end{acknowledgments}

\bibliography{main}
\bibliographystyle{apsrev4-2}

\end{document}